\documentclass[%
reprint,
superscriptaddress,
amsmath,amssymb,
aps,
prb,
]{revtex4-2}

\usepackage{xcolor}
\usepackage{graphicx}
\usepackage{dcolumn}
\usepackage{bm}
\usepackage{mathrsfs}
\usepackage{hyperref}
\usepackage[normalem]{ulem}
\usepackage{lipsum}
\usepackage{IEEEtrantools}
\usepackage[version=3]{mhchem}
\usepackage{verbatim}
\usepackage{siunitx}
\usepackage{enumitem}
\usepackage{textcomp}
\bibpunct{[}{]}{;}{n}{}{}

\setlist[itemize]{noitemsep, topsep=0pt}
\begin{document}

\title{On the Uncertainty Estimates of \\ Equivariant-Neural-Network-Ensembles Interatomic Potentials}

\author{Shuaihua Lu}
\affiliation{Key Laboratory of Quantum Materials and Devices of Ministry of Education, School of Physics, Southeast University, Nanjing 21189, China}
\affiliation{The NOMAD Laboratory at the Fritz Haber Institute of the Max-Planck-Gesellschaft and IRIS Adlershof of the Humboldt-Universit\"{a}t zu Berlin, Germany}
\author{Luca M. Ghiringhelli}
\email{luca.ghiringhelli@physik.hu-berlin.de}
\affiliation{Physics Department and IRIS Adlershof of the Humboldt-Universität zu Berlin, Germany}
\affiliation{The NOMAD Laboratory at the Fritz Haber Institute of the Max-Planck-Gesellschaft and IRIS Adlershof of the Humboldt-Universit\"{a}t zu Berlin, Germany}
\author{Christian Carbogno}
\affiliation{The NOMAD Laboratory at the Fritz Haber Institute of the Max-Planck-Gesellschaft and IRIS Adlershof of the Humboldt-Universit\"{a}t zu Berlin, Germany}
\author{Jinlan Wang}
\affiliation{Key Laboratory of Quantum Materials and Devices of Ministry of Education, School of Physics, Southeast University, Nanjing 21189, China}
\author{Matthias Scheffler}
\affiliation{The NOMAD Laboratory at the Fritz Haber Institute of the Max-Planck-Gesellschaft and IRIS Adlershof of the Humboldt-Universit\"{a}t zu Berlin, Germany}

\date{\today}

\begin{abstract}
Machine-learning (ML) interatomic potentials (IPs) trained on first-principles datasets are becoming increasingly popular since they promise to treat larger system sizes and longer time scales, compared to the {\em ab initio} techniques producing the training data. 
Estimating the accuracy of MLIPs and reliably detecting when predictions become inaccurate is key for enabling their unfailing usage. 
In this paper, we explore this aspect for a specific class of MLIPs, the equivariant-neural-network (ENN) IPs using the ensemble technique for quantifying their prediction uncertainties. We critically examine the robustness of uncertainties when the ENN ensemble IP (ENNE-IP) is applied to the realistic and physically relevant scenario of predicting local-minima structures in the configurational space.
The ENNE-IP is trained on data for liquid silicon, created by density-functional theory (DFT) with the generalized gradient approximation (GGA) for the exchange-correlation functional. Then, the ensemble-derived uncertainties are compared with the actual errors (comparing the results of the ENNE-IP with those of the underlying DFT-GGA theory) for various test sets, including liquid silicon at different temperatures and out-of-training-domain data such as solid phases with and without point defects as well as surfaces. 
Our study reveals that the predicted uncertainties are generally overconfident and hold little quantitative predictive power for the actual errors.
\end{abstract}

\maketitle

\section{Introduction}
Machine-learning (ML) surrogate models for interatomic potentials (IPs) trained on {\em ab initio} reference data are playing an increasingly important role in condensed-matter simulations. Such MLIPs promise to achieve a converged treatment of the statistical mechanics of the modeled atomistic systems, because they can address both larger system sizes and longer timescales than the computationally much more costly {\em ab initio} calculations. Obviously, this also requires the MLIPs to retain an accuracy that is at least comparable to that of the reference data~\cite{xie2018crystal,deringer2019machine,unke2021machine} 
To this end, many advanced frameworks~\cite{chmiela2017machine,schutt2018schnet,unke2021machine,sauceda2022bigdml,kocer2022neural} have been proposed and employed to study a wide range of physical mechanisms, such as phase transformations~\cite{cheng2020evidence, deringer2021origins,kapil2022first} as well as thermal transport in crystals~\cite{Sosso.2012,Korotaev:2019ep,Verdi.2021,Langer2023}, chemical reaction processes \cite{timmermann2020iro, li2022origin, li2022smallest}, and more. 

Accordingly, the basic motivation behind MLIPs is essentially the same as that behind classical force fields and semi-empirical interatomic potentials \cite{brooks1983charmm,daw1984embedded,weiner1984new,tersoff1988new,brenner1990empirical,van2001reaxff,oostenbrink2004biomolecular}. Those are based on physically and chemically motivated equations and therefore offer less flexibility than MLIPs. 
MLIPs and force fiels share the same fundamental key problem:
What is their reliability when used for situations that are different to those used for training? 
An answer to this questions require to establish (i)~quantitative uncertainty estimates as well as a (ii)~metric that measures the ``distance'' between prediction and training data points.
To this end, one may determine the domain of applicability of MLIPs \cite{sutton2020identifying},~i.e.,~the domain outside of which MLIPs will likely fail. However, this is challenging and hardly ever done.

In this paper, we employ a more pragmatic route: Instead of determining the applicability domain via data-science methods, we explore the applicability of MLIPs and their uncertainty estimates for distinct, physically motivated applications. Although frequently used terms such as ``different to the training data set'', ``out-of-training domain'', and ``out-of-distribution data'' are not well defined in this case, this is still useful, since it facilitates the physical interpretation of the results. Accordingly, we here focus on the reliability of uncertainty estimates obtained by MLIPs for different applications. Obviously, if reliable uncertainty estimates existed, they would also enable the identification of the domain of applicability. 

Improving this nebulous situation is a pressing issue in the field, since thermodynamic or kinetic simulations are guaranteed to explore also uncommon regions of configurational space that are likely not covered by the training data set. An example may be that a rare event that substantially influences the dynamics may be missed by the MLIP. For instance, the rare, but spontaneous formation of defects is know to be a key trigger for phase transition
and for limiting heat transport~\cite{Knoop2023}. Were the critical defect or the new phase not known beforehand, they will not be part of the training data, so that it is largely unclear if the MLIP would be able to find it at all. Indeed, rare events of (so far) unknown nature are key for many properties described by the statistical mechanics of materials.
Another example for this ``out-of-training domain'' dilemma is crystal-structure prediction, in which, by definition, a global 
configurational search is performed to discover structural local minima that are unknown during the training process. For these kind of applications, it is essential that the surrogate models yields accurate predictions over the whole configurational space, or, at least 
can reliably quantify the expected confidence/uncertainty for every predicted configuration \cite{lakshminarayanan2017simple}. 
Let us emphasize that this question is also pivotal to exploit active-learning strategies,~e.g.,~in the use cases described above. Reliable uncertainty estimates are key to efficiently advance and improve MLIPs in these approaches. Otherwise, the active learning may just proceed in the wrong direction and, e.g., reinforce the bias imposed by the initial training data. 

Uncertainty quantification for ML models is model-class specific task. For instance, Gaussian potential (GP) models\cite{bartok2010gaussian}, which are based on Bayesian inference, provide statistical uncertainty using posterior predictive variances by construction. However, the reliability of GPs' built-in uncertainty estimates degrades in the ``out-of-training domain'' \cite{bartok2022improved}. 
A neural-network (NN) MLIP does not {\em inherently} provide prediction variances, but a properly created ensemble of NNs will yield a distribution of predictions for the target properties (for an MLIP, energy and forces) and the variance of this distribution can be used as a proxy for the uncertainty estimate.
Such ensemble may be created by varying (subsampling) the training data \cite{musil2019fast},  the NN structure\cite{musil2019fast, jeong2020efficient}, the initialization of the NN training \cite{gastegger2017machine,zhang2018deep}, or combinations thereof. 

Naturally, the use of ensembles increases the computational workload for predictions considerably compared to a single NN. Accordingly, several recent works have proposed strategies and metrics to aid, improve, and accelerate uncertainty estimates for NN MLIPs \cite{liu2018molecular,janet2019quantitative,hu2022robust,zhu2023fast,carrete2023deep,wollschlager2023uncertainty}.
These metrics typically show good performance, especially in the context of active learning. Indeed, more extended and systematic comparisons of several uncertainty-estimate approaches tested on a broad range of molecular, bulk, and surface systems have shown \cite{hirschfeld2020uncertainty,kahle2022quality,tan2023single} that the accuracy of these different methods is very similar to each other, but slighlty lower than ensemble methods, which show the highest reliability. Furthemore, these studies revealed that all methods tend to underestimate the uncertainties (i.e., they are overconfident). 
We note that the thereto used ``out-of-domain'' data is much more closely related to the training data than the physical test sets used in this work. Furthermore, a linear correlation between uncertainty estimates and actual errors was found in these works~\cite{kahle2022quality,tan2023single}, suggesting that reliable error predictions for NN MLIPs can straighforwardly be obtained by calibrating the uncertainty estimates.

In this work, we go beyond the previous studies and explicitly explore the reliability of uncertainties for regions in configuration space that are not necessarily well covered by the actual training data.  
As discussed above, these test sets are motivated by physical knowledge for a well-known system, elemental silicon. In practice, we thus construct an MLIP for elemental silicon by training an NN ensemble (NNE) on liquid Si, i.e., on the energies and forces observed in high-temperature {\em ab initio} MD simulations in the canonical ensemble over a representative range of pressures viz.~densities. This ensures that the MLIP training data contains a quite diverse set of configurations, since atoms in the liquid phase visit many local geometries, explore different short-range order and bonding situations. From a physical point of view, it is hence reasonable to assume that the training data largely also covers the physical and chemical situations found at lower temperatures,~i.e,~in different bulk solids, defects, and surfaces.
Next, we benchmarked the NNE MLIP on the eight known crystal phases of Si \cite{PhysRevX.8.041048}, as well as on important intrinsic point defects and on the reconstruction of the Si (100) surface. 
In particular, we analyze the uncertainty estimates on (formation) energies of these structures. The latter are compared to the uncertainty estimates of configurations in the liquid phase at different temperatures and to the difference between the NNE predictions and the ground-truth, henceforth called {\em actual errors}.
Our results show that the created MLIPs provide a reasonably good qualitative description of the tested systems. However, quantitative values, e.g., on defects formation energies, are not reliable, and this may well have dramatic consequences for numerous material's properties. 
Crucially, the uncertainty estimates as derived from the variance of the ensemble's distributions are not able to reliably predict the actual errors.
For out-of-domain test data, one could not even use the uncertainty estimates for a trusted warning if the results obtained with such MLIPs are just slightly quantitatively inaccurate or qualitatively wrong.

The structure of this paper is as follows: in section \ref{sec:construction} equivariant neural network ensembles (ENNE), i.e., the model class for MLIPs discussed in this paper, are introduced together with the adopted training and validation protocol. 
We then report on the testing of the trained ENNE-IP on liquid configurations not used for training (section \ref{sec:liquid}), the equations of state of the eight known solid phases of silicon (section \ref{sec:eos}), and defected structures (section \ref{sec:defects}). Next, in section \ref{sec:errors}, the reliability of the uncertainty estimates for the trained ENNE-IP is quantitatively analyzed.  

\section{Results}
\subsection{Construction of Neural-Network-Ensemble Interatomic Potentials} 
\label{sec:construction}
\begin{figure*}[ht]
    \centering
    \includegraphics[width=0.8\textwidth]{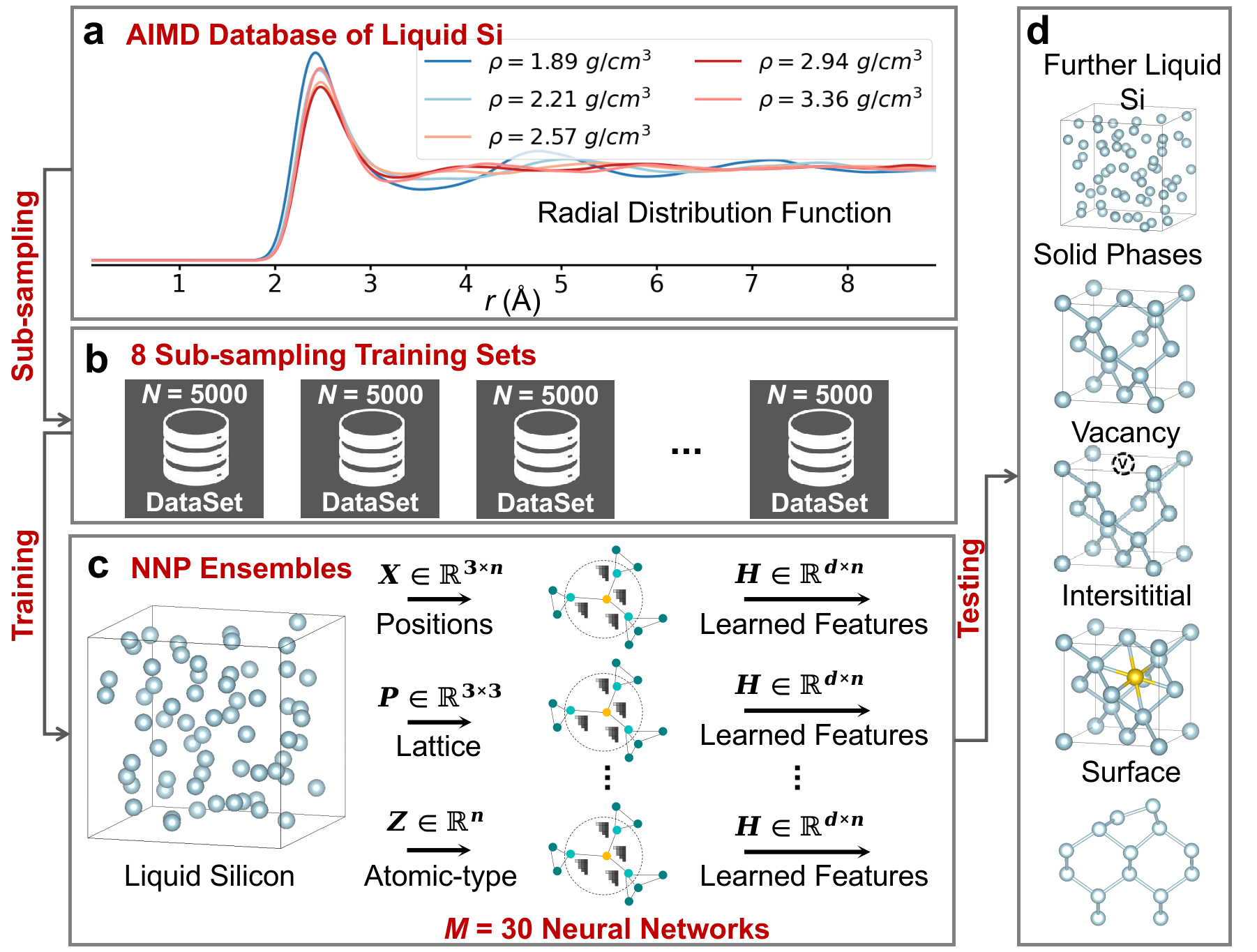}
    \caption{Workflow of the creation of NN-ensemble MLIPs, i.e. of the overall training and testing process. (a) Radial distribution functions of AIMD trajectories simulated at 5 different densities and $T= 2\,000$~K. 
    (b) Randomly sub-sampled training sets with different initialization of the NN parameters. 
    (c) Construction of the NN-ensemble MLIP.  $n$ and $d$ refer to the number of atoms in one crystal structure and the dimension of the learned features from the neural network, respectively. (d) Systems used for predictions and uncertainty estimates, including liquid silicon at different temperatures and “out-of-training-domain” data such as solid phases with and without point defects as well as surfaces.}
    \label{fig:workflow}
\end{figure*}

Figure \ref{fig:workflow} summarizes the workflow for training the ENNE-IPs. The initial step involves the preparation of the data sets that are used for training. 
For an MLIP to be both accurate and applicable to new and not-yet known situations, it is crucial that the training set contains ample information regarding the possible local atomic environments and bonding situations. 
As the atoms in a liquid visit a huge variety of local geometries and interatomic interactions, we create the training data set by {\em ab initio} MD (AIMD) simulations of liquid Si. The many-electron exchange and correlation effects are approximated by the generalized gradient approximation (GGA), specifically, the PW91 exchange-correlation\cite{PW91}
We find that it is important that the training data set covers liquids at different densities: If the NN MLIP is trained solely on liquid silicon at one density, we obtain a systematic error when predicting the energies of liquid silicon at other densities. Thus, we consider data from liquid Si at different densities. As the number of densities of the liquid-phase data increases, the error decreases, reaching convergence at five densities (as demonstrated in Fig. S1 in the electronic supplementary information, ESI). The initial density for the AIMD of liquid Si was chosen as the its experimental density at ambient pressure and a temperature of $1\,687$~K \cite{remsing2017dependence}. The temperature of the simulations used for training was set to $2\,000$~K, which is well above the experimental melting temperature. 
The other densities (or pressures) were chosen in the vicinity of this experimental density. The choice of the training densities is dictated solely by the need to sample physically relevant configurations. 

In this work, the MLIP model class is the E(3) equivariant neural network (ENN) of Batzer {\em et al.} \cite{batzner20223}, as implemented in the NequIP software.
This approach has been used on a challenging and diverse set of molecules and materials, achieving state-of-the-art accuracy. 
As represented in Fig. \ref{fig:workflow}c, the MLIP is an ensemble of NNs, consisting of $M$ individual neural networks ($M$ refers to the size of the ensemble) with the same architecture but different initial conditions (i.e., different randomly initialized sets of nodes' weights). The {\em actual error} $ \epsilon_{i} $ is defined as the difference between the average predicted value of the ENN ensemble (ENNE) IP and the DFT-GGA ground-truth value. As estimator of the {\em model uncertainty} for a given test configuration $i$, we use the standard deviation $\sigma_{i}$~\cite{zhang2018deep} of the predictions by the trained ENNs in the ensemble. The next crucial step was to determine the minimal value of $M$ for which the {\em model uncertainty} is converged. Previous studies have utilized ensembles composed of a limited number of neural networks, with $M\leq 10$, and the dependence between the ensemble size and the level of model uncertainty has not been extensively studied \cite{gastegger2017machine,zhang2018deep,kahle2022quality}. 
We found (see Fig. S3 in the ESI) that $\sigma$ steeply increases with the ensemble size and a plateau value is reached around $M=30$.
Hence, we chose to use ensembles of this size in all of our experiments. It is important to note that, at least for the system investigated in this work, the use of $M=30$ individual neural networks is necessary for obtaining converged uncertainty estimates. Using $M=30$ in actual calculations is, however, expensive. Interestingly, we notice that the plateau value for $\sigma$ is in all studied cases within a multiplicative factor 2--2.5 from the value obtained by an ensemble with only 4 ENNs. This means that, if $\sigma$ turned out to be a good predictor for $\epsilon$ in general, one might well use the unconverged average value obtained  with few NNs and scale it up. Further details on the data generation, data set construction, and the ensemble model are presented in the ``Methods'' section.

Our assessment of the converged size of the ensemble is based on the strategy where the training set is fixed and the members of the ensemble are trained starting from different random initial guesses of the NN parameters. We have looked into the Pearson pair correlation of the errors over a hold-out (test) set of $1\,000$ configurations in the liquid phase. In practice, the prediction of each NN in the ensemble generates an array where each component is the error $\epsilon_i$ for one test configuration and the pair correlation between any two such arrays was recorded. A high Pearson correlation among these arrays signals that the NNs in the ensemble, despite having different values of the training parameters, essentially make the same kind of error on unseen data points. We found (Fig. S10 in the ESI) that within the ensemble the Pearson correlation is as high as $0.85 \pm 0.04$. Aiming at generating an ensemble with less linearly correlated errors, we then trained eight ensemble models, each on a different training set of $5\,000$ configurations, extracted from a training pool of $25\,000$ configurations in the liquid phase. Each ensemble was composed of $N=30$ members following the different-initial-guess approach. However, we found (Fig. S10 in the ESI) that also within this type of ensemble the Pearson correlation of the errors, over the same test set as defined above, was at least as high as among the 30 models in each ensemble ($0.88 \pm 0.05$). In the following, we therefore proceed analyzing the results for only one ensemble, i.e., one training set and 30 models trained with different initial conditions.

\subsection{Liquid silicon} 
\label{sec:liquid}
\begin{figure}[ht]
    \centering
    \includegraphics[width=\columnwidth]{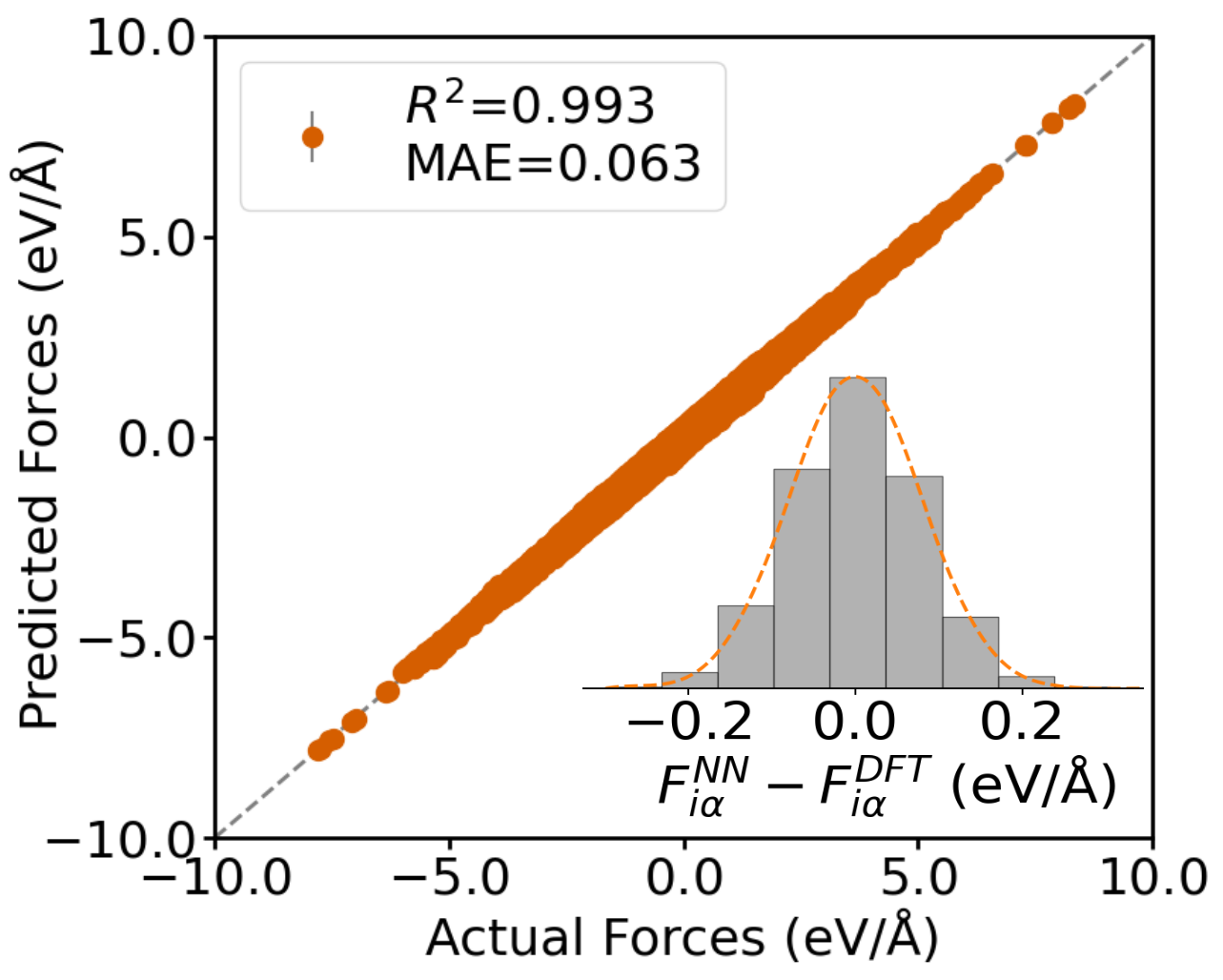}
    \caption{The prediction of force components using the ENNE-IP against the DFT-GGA-calculated forces for liquid-Si configurations not used for training, from DFT-GGA trajectories.
    }
    \label{fig:MD}
\end{figure}

The performance of the ENNE-IP was at first evaluated by predicting the forces in liquid-Si configurations not used for training, extracted from portion of the AIMD trajectories outside the intervals from which the training data had been extracted.
The parity plot for the force components, values predicted by the ENNE-IP vs actual DFT-GGA values are 
depicted in Fig. \ref{fig:MD}. The predictive accuracy was evaluated based on the mean absolute error (MAE) of the force components. The comparison between the DFT-GGA calculations and the ENNE-IP's predictions showed an MAE of 0.063 eV/\AA\ for the forces. 
The results reflect that the ENNE-IP accurately captures the chemical and structural features of liquid silicon at the temperature and pressure conditions used for the training. 

We now generalize the situation and analyze the uncertainty estimates of the ENNE-IP for various other systems, including liquid silicon at different temperatures and solid phases, with and without point defects, as well as surfaces (Fig. \ref{fig:workflow}d). 

\begin{figure}[ht]
    \centering
    \includegraphics[width=\columnwidth]{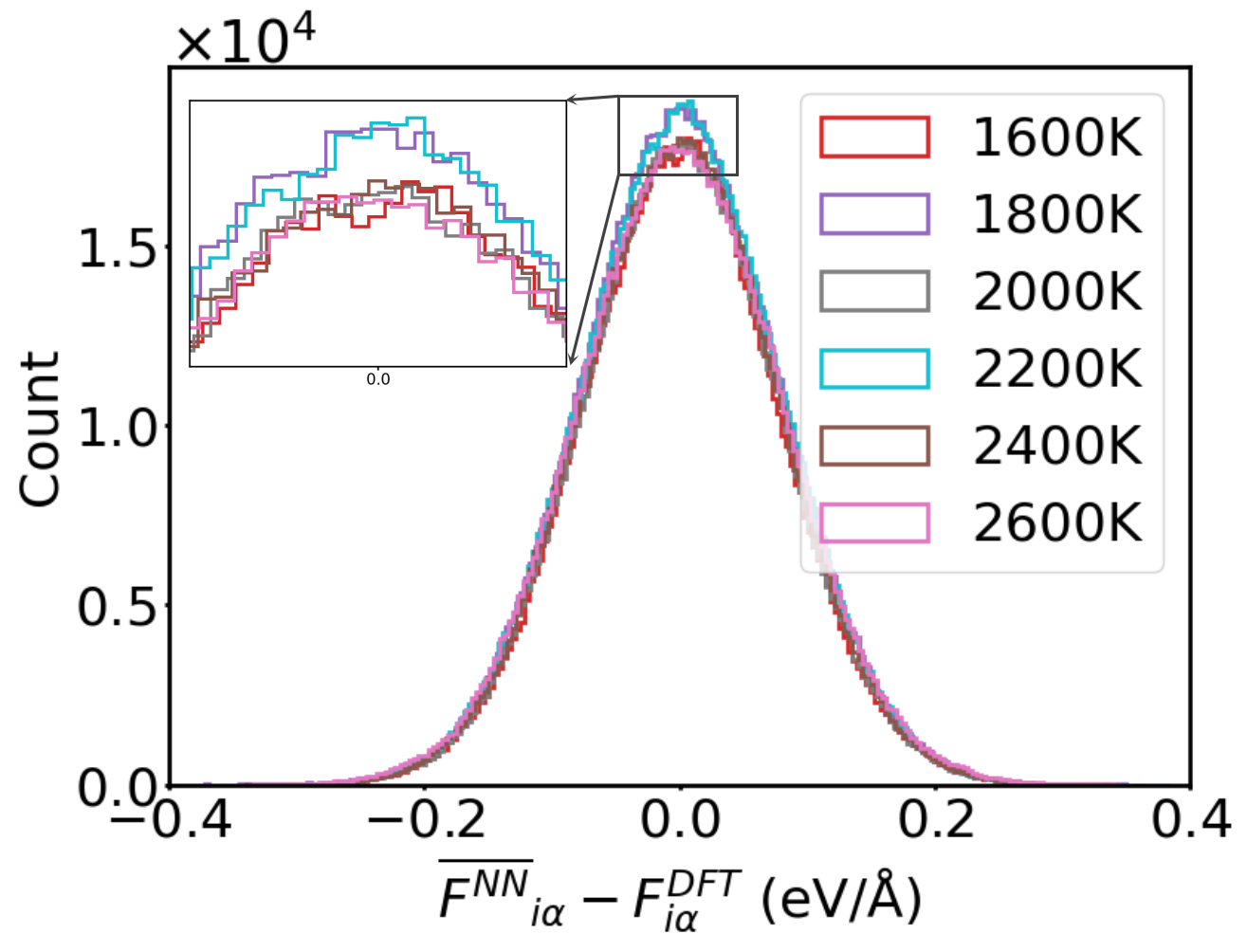}
    \caption{Histograms of the actual error of force components reveal an approximately Gaussian distributed error.
    }
    \label{fig: MD prediction at the different temperature}
\end{figure}

Our first prediction examples on data not used for training are still in the liquid phase of silicon, at different temperatures. 
As mentioned earlier, the training data correspond to $T=2\,000 $~K, and now we study results for temperatures in the range from $1\,600$ to $2\,600$~K. A histogram of the actual error of force components for each temperature is shown in Fig. \ref{fig: MD prediction at the different temperature}.
We observe that the distribution of actual errors is essentially independent of the temperature difference between the liquid silicon configurations in the predictions set and training set increases.

\begin{figure*}[ht]
    \centering
    \includegraphics[width=\textwidth]{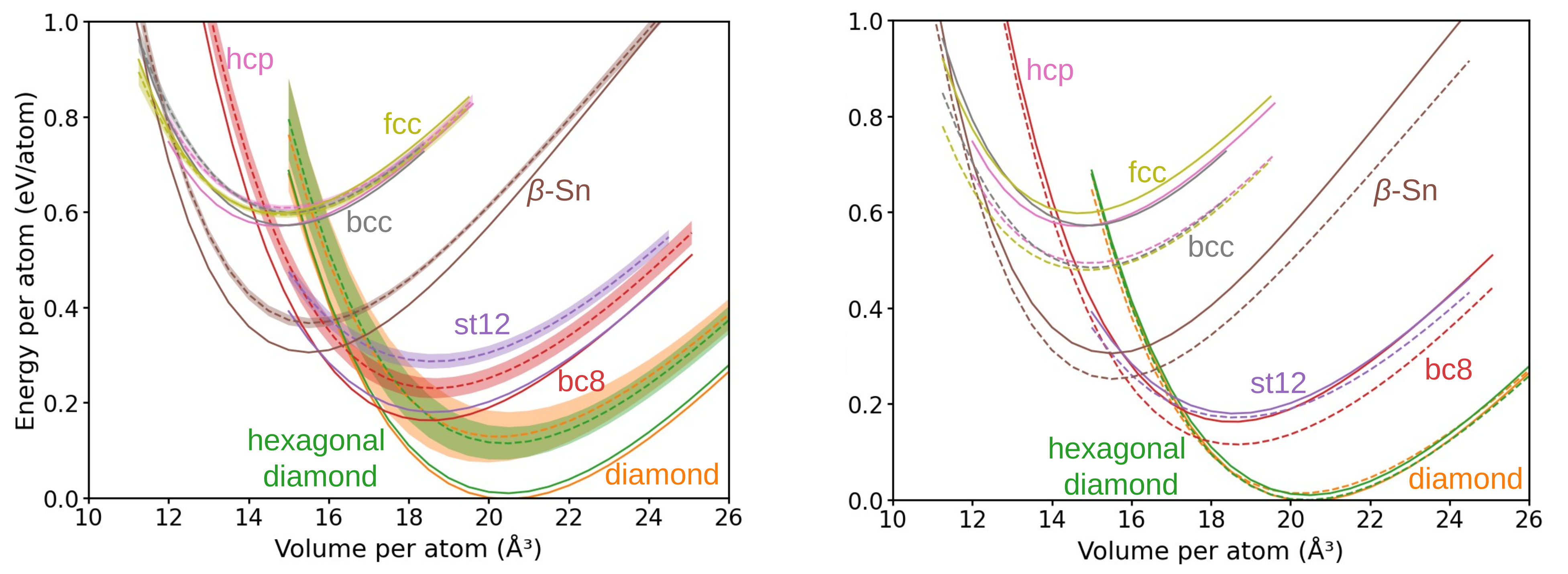}
    \caption{Comparison of the equation of states of the ground-truth DFT-GGA calculations (full lines) and the predictions by the ENNE-IP (dashed lines). Left panel: The energies predicted by the ENNE-IP are directly compared to the DFT-GGA energies and the energy zero is set to the DFT-GGA ground-state configuration (minimum of the cubic diamond structure). The colored band represent the uncertainty estimates for the ENNE-IP predictions. Right Panel: as a guide for the eye, the ENNE-IP-predicted values are shifted 0.115 eV/atom downwards, so that the zero energy is set independently to the ground-state configuration for both ENNE-IP (minimum of the hexagonal diamond structure) and DFT-GGA (same as left panel). 
    }
    \label{fig:equation-of-state}
\end{figure*}

\begin{figure}[t]
    \centering
    \includegraphics[width=0.8\columnwidth]{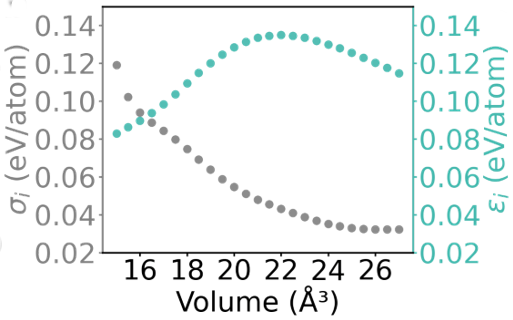}
    \caption{Comparison of the predicted uncertainty $\sigma$ and the actual errors $\epsilon$ for the cubic diamond phase over the studied volumes.
    }
    \label{fig: sigma-epsilon}
\end{figure}

\begin{figure}[ht]
    \centering
    \includegraphics[width=\columnwidth]{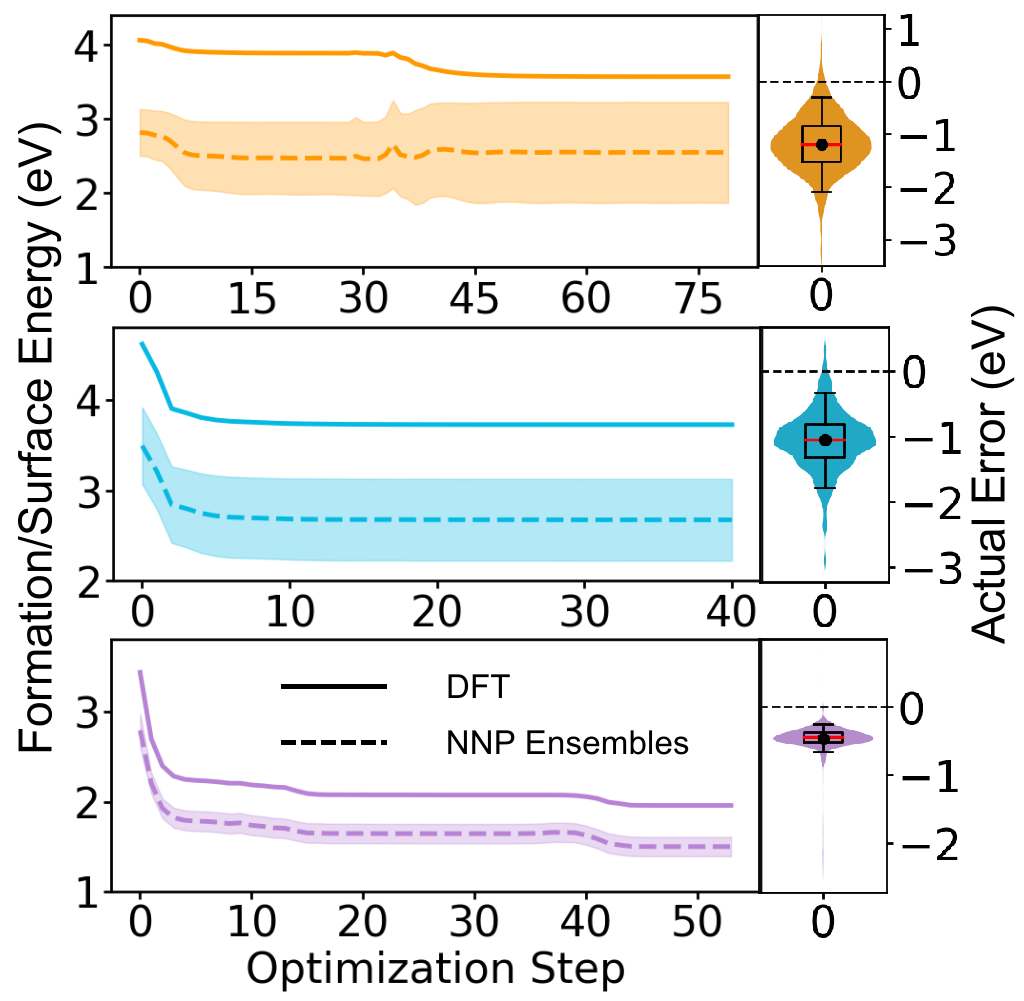}
    \caption{Test of the ENNE-IP on (a) Si Vacancy, (b) Si interstitial and (c) Si(100) Surface. The solid line is the trajectory of geometry optimization computed using the DFT (solid lines) from the unrelaxed crystal structure to the relaxed one. The dashed line is the trajectory of geometry optimization computed using the ENNE-IP. The color area depicts the uncertainty $\sigma$. The {\em violin plots} show the distribution of actual errors. The black dot represents the mean actual error, the orange lines are the median actual error, the boxes are the quartiles, and the whiskers are the 5\% and 95\% actual error}
    \label{fig: Extrapolation to Si Vacancy, Self-Interstitial and Si(100) Surface}
\end{figure}

\subsection{Equations of state for 8 crystal phases of Si}
\label{sec:eos}
Figure \ref{fig:equation-of-state} displays the calculated equation of states for 8 crystal structures of silicon: the experimental ground-state diamond structure, the closely related hexagonal diamond, and the high-pressure structures $\beta$-Sn, bc8, and st12, hexagonal close-packed (hcp), body-centered cubic (bcc), and face-centered cubic (fcc) \cite{PhysRevX.8.041048}. The full lines are the DFT-GGA results and the dashed lines are the results of our ENNE-IP. To facilitate the qualitative comparison, in the right panel of Fig. \ref{fig:equation-of-state}, the energy zeros are set independently for the ENNE-IP and DFT-GGA results, so that the respective ground-state configurations have zero energy. This implies a relative downshift of all ENNE-IP results by 0.115 eV/atom compared to the ground-truth DFT-GGA results. At a first glance, one may conclude that the qualitative description of both treatments is essentially the same. Obviously, quantity can well affect the quality, and we see, for example, that the ENNE-IP predicts that the hexagonal-diamond structure is the most stable one, which is not correct.  The phase transition between the diamond and the $\beta$-tin structure was compared by means of the Gibbs construction (common tangent of the E(V) curves between the two phases, see Fig. S5 in the ESI). For ENNE-IP, a value of $12.0 \pm 3.1$~GPa was found, to be compared with 10.4 for DFT-GGA, thus in good agreement.
The detailed results for all studied structures, including the equilibrium volume, the depth of the energy minimum (cohesive energy relative to the ground state), and bulk modulus (related to the curvature of the $E(V)$ curve) are shown in Fig. S5 in the ESI. 

Let us now look at the uncertainty estimates, if they would at least give a warning about the reliability of the results.
The left panel of Fig. \ref{fig:equation-of-state} shows the uncertainty estimates $\sigma$ (the colored bands) for the  $E(V)$ curves of each crystal structure calculated with ENNE-IP.

\begin{figure*}[t]
    \centering
    \includegraphics[width=0.9\textwidth]{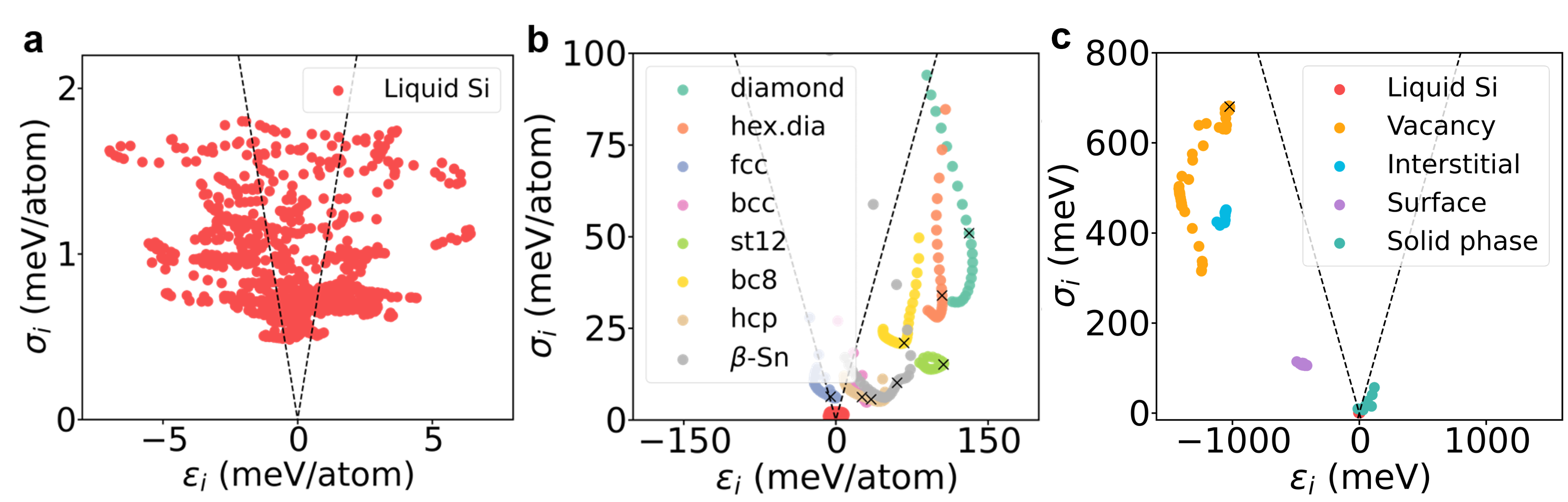}
    \caption{The ensemble uncertainty $ \sigma$ is plotted against the actual error $ \epsilon$, calculated as the difference between the predicted value and the DFT-GGA calculated value for five test sets, including (a) liquid silicon, (b) bulk crystal structures, (c) Si vacancy, Si interstitial atom, and Si(100) surface. The dashed line represents the ideal relationship between the actual error and uncertainty, $\sigma = |\epsilon|$. The crosses mark the minimum energy structure (according to ENNE-IP) in the equations of state (panel b) or geometry optimization (panel c). In panel c, the energies are normalized, i.e., energy per atom for the bulk phases (as in panels a and b), formation energy per defect for the defected phases, and surface energies per surface atom for the surface. 
    }
    \label{fig: actual error vs. Standard Deviation}
\end{figure*}

For reliable uncertainties, the colored bands should contain the ground-truth values (solid lines). Even though the larger the actual error (the vertical distance between the solid and dahsed lines for each phase) the larger the predicted uncertainties tend to be, the latter are still overconfident, i.e., the ground-truth value is outside the range of the predicted uncertainties for most of the volumes in the equations of state, for all phases.

To further analyze the relationship between estimated uncertainty and actual error, we have plotted for the cubic-diamond phase these two quantities at all investigated volumes along the equation of state (Fig. \ref{fig: sigma-epsilon}). Not only the desired behavior of having the actual error $\epsilon$ smaller than the uncertainty $\sigma$ is verified only in a small interval, but, strikingly, overall $\sigma$ is mostly not related to $\epsilon$, in particular, not monotonically. In other words, even the practical solution of multiplying the uncertainty by a factor in order to try and always bound the actual error, appears to be unhelpful.

\subsection{Defect Phase and Surface}
\label{sec:defects}
The ENNE-IP is applied to more complex cases, the solid phases with point defects as well as surfaces, including diamond vacancy \cite{estreicher2011activation}, hexagonal diamond interstitial \cite{PhysRevLett.83.2351} and Si(100) surface\cite{PhysRevLett.43.43, ramstad1995theoretical}. 
A rich complexity of bonding emerges in vacancy/interstitial and on surfaces due to the subtle interplay of strain effects with the chemistry of dangling bonds. This complexity makes formation energies (surface energy), and particularly the energies and geometries of various reconstructions, a sensitive test of the accuracy of interatomic potential.

The formation energies of a vacancy and a hexagonal interstitial in Si, as well as the surface energy for the Si(100) surface, were predicted using ENNE-IP. The configurations in geometry optimizations from DFT-GGA calculations were used as the test sets. It is worth noting that the considered structures of vacancy and interstitial are both ${5 \times 5 \times 5}$ supercell of the eight-atom diamond cubic cell, and that of the surface is ${2 \times 1 \times 10}$ supercell. These supercell sizes were found to be the converged ones for the DFT-GGA calculations. As shown in panels a and b of Fig. \ref{fig: Extrapolation to Si Vacancy, Self-Interstitial and Si(100) Surface},
both the actual error and uncertainty in the formation energies of the vacancy and interstitial were substantial, around 1 eV. 
Moreover, the results showed that the actual value of the formation energy for these vacancies  is never included within the uncertainty interval. 
Fig. \ref{fig: Extrapolation to Si Vacancy, Self-Interstitial and Si(100) Surface}c shows the surface energy of the tilted-dimer ${2 \times 1}$ reconstruction, one of the low-energy configurations of the (100) surface, which forms spontaneously from the as-cut surface. ENNE-IP captures the tilted-dimer geometry of the reconstructed surface in reasonable agreement with the DFT-GGA calculations, but the predicted formation energy is 0.5~eV lower than the DFT-GGA value and, again the uncertainty interval does not include the DFT-GGA value.

In Fig. \ref{fig: Extrapolation to Si Vacancy, Self-Interstitial and Si(100) Surface}, we also present the distribution of actual errors from the ensemble. 
The results reveal that although some individual models have actual errors that cross the line where the error equals zero, the mean actual error is still far from this line. This suggests that the uncertainty may underestimate the actual error. To further explore this issue, we also conducted an analysis of the actual errors for various supercell sizes, as presented in Fig. S6-S8 in the ESI. The results show that the size of the supercell has little effect on both the uncertainty and actual error.

\subsection{Predicted Uncertainty vs Actual Error}
\label{sec:errors}
We calculated the actual error and uncertainty from the ensemble (30 ENN models)
for all configurations in the test sets, including liquid silicon at $2\,000$ ~K, solid phases, Si vacancies, Si interstitial, and surface Si(100). This step aims to further understand the connection between uncertainty and actual error. The relationship between the actual error and uncertainty is displayed in Fig. \ref{fig: actual error vs. Standard Deviation}. Fig. \ref{fig: actual error vs. Standard Deviation}a shows the actual error vs uncertainty for liquid silicon at various temperatures. For the uncertainty $\sigma$ to be considered a good predictor for the actual error $\epsilon$ or at least an upper bound, the data points should lie around the dotted lines ($\sigma = |\epsilon|$) or mostly included within these lines.
For liquid Si, the data points reasonably fall near or inside the dotted lines. One should notice also the very low values for both quantities, as expected, since the ENNs were trained on liquid silicon.  Still, ``large'' values of $\sigma$ do not necessarily imply large errors~$\epsilon$,~e.g.,~several configurations exhibit vanishing error despite having a $\sigma > 1$~meV/atom. Neither does a particularly low value of $\sigma$ guarantee high-accuracy predictions, e.g.,~the majority of  configurations with $\sigma < 1$~meV/atom do not fall within the $\sigma = |\epsilon |$ cone.  

The result for the eight solid phases is shown in Fig. \ref{fig: actual error vs. Standard Deviation}b, where all the points along the equations of state are reported. Both $\sigma$ and $\epsilon$ are much larger than for the liquid case and the data points fall almost all in the region $\epsilon > \sigma$, up to $\epsilon \sim 5\sigma$, varying depending on the type of crystal phase and the volume. Moreover, in particular focusing on the diamond, hexagonal diamond, and bc8 phases, the error is essentially unrelated to the uncertainty, as the uncertainty spans a broad range of values when the volume is changed, while the error remains (large and) almost constant. In Fig. \ref{fig: actual error vs. Standard Deviation}c, we also include the defect phases. Both actual errors and predicted uncertainty are even larger and although the two quantities roughly grow concurrently, there is hardly any useful relationship between them. For instance, in the case of the relaxation trajectory of the vacancy, where configurations are rather similar to each other, we observe a huge change in predicted uncertainty (more than a factor 2) with the actual error staying essentially constant around 1 eV.

We also checked the validity of these observations at various sizes of the training data set and we found (see Fig. S9 in the ESI) that the lack of useful relationship between $\sigma$ and $\epsilon$ holds at all dataset sizes, between 500 and $5\,000$ data points.

\section{Discussion}
Estimates for prediction uncertainty are important for determining when MLIPs can be trusted. We constructed a set equivariant-neural-network-ensemble IPs on the liquid silicon and compared the uncertainty with actual errors for various test sets, including liquid silicon at different temperatures and out-of-training-domain data such as solid phases with and without point defects as well as surfaces. We use the standard deviation of the predicted values from the ensemble as an uncertainty estimate. This popular approach to uncertainty calculation has multiple advantages, including being simple to implement and parallelize, and maintaining desirable features of the original regression model.

However, by analyzing a test set constructed to mimic a crystal-structure search scenario, for a system that is trained with configurations in the liquid phase, we found that the predicted uncertainty is a poor predictor for the actual error. Not only it is mostly overconfident (uncertainty lower or much lower than the actual error) but both quantity are hardly related. Indeed, we have analysed a few paths in the configurational space where the two quantities change in opposite directions (uncertainty decreases while error increases) or on varies and the other stays constant. This aspect would hamper any attempt to calibrate the uncertainties to create an upper bound for the predicted errors,~e.g.,~by just scaling the predicted uncertainty by a constant factor. In fact, there could always be a region of the configurational space where the error would be even larger than such amplified uncertainty, yet containing relevant, metastable structure, such as technologically important defect structures. 

Eventually, let us note that it is obviously possible to choose as an upper bound for the trustworthy region an uncertainty $\hat{\sigma}$ equal to the maximum uncertainty observed in the liquid-phase test data
($\sim 2$~meV/atom, see Fig. \ref{fig: sigma-epsilon}) 
The ENN would then be retrained whenever larger uncertainties are met. However, this largely defeats the original intention, since it would require retraining in all the discussed cases, no matter if actual qualitative or quantitative agreement is already achieved. For instance, the uncertainties for the configurations used for both the fcc and hcp equations of state have similar values (i.e., around 2 mev/atom, see Figs \ref{fig:equation-of-state}, left panel, and \ref{fig: actual error vs. Standard Deviation}). However, on the one hand, the fcc equation of state predicted by the ENNE-IP is quite accurate, i.e., the uncertainty band mostly overlap with the ground truth in Fig. \ref{fig:equation-of-state} and the $(\sigma,\epsilon)$ points are very close to the $\sigma = - \epsilon$ line in Fig. \ref{fig: actual error vs. Standard Deviation}. On the other hand, the prediction on the hcp equation of state is much less accurate, even from a qualitative point of view, i.e., it is predicted to be less stable than fcc, contrary to the ground truth.
In other words, such a strategy would inherently imply that the here discussed MLIPs are only applicable to the system/conditions they have been trained on. This is obviously not true, as the above fcc example showcases. However, the here discussed ensemble-uncertainties provide only very limited guidance in disentangling trustworthy from unreliable regions.

Further research is needed on how to properly calibrate ensemble MLIPs, and an especially interesting avenue is, in our opinion, to explore whether there is an accuracy-confidence trade-off. In some other ML applications instead of MLIPs, the distance to available data in the latent space of a NN model is proposed to control actual error in chemical discovery \cite{janet2019quantitative,hirschfeld2020uncertainty,busk2021calibrated}. These works provide calibrations to improve uncertainty estimates from feature distances or NN-predicted variances. However, they do not demonstrate that the approaches can be used more generally or propose directly calibrating ensemble errors. Domain of applicability (DA) \cite{sutton2020identifying} is another potential method to calibrate current uncertainty estimates. An extension of the original DA work could provide a descriptive characterization, in terms of inequalities on simple local-order descriptors, of regions of the input space where the error is expected to be particularly large/low, thus offering the possibility to dynamically fine tune the numerical precision of the MLIP (e.g., in terms of number of basis functions).

\section*{Data availability}
All data can be found on the NOMAD archive (\url{https://doi.org/10.17172/NOMAD/2023.08.25-1})

\section*{Code availability}
 FHI-vibes was used to generate all data and analysis in the paper and are freely available online in the cited publication. All electronic structure calculations were done using FHI-aims, which is freely available for use for academic use (with a voluntary donation) (\url{https://fhi-aims.org/get-the-code-menu/get-the-code}). NequIP was used to construct interatomic potential, and a tutorial notebook can be found on the NOMAD AI Toolkit \url{https://gitlab.mpcdf.mpg.de/nomad-lab/ai-toolkit/tutorial-nequip}

\section*{Acknowledgements}
We acknowledge David Villareal for insightful discussions in the design and preliminary steps of this project.
This work was funded by the National Key Research and Development Program of China (2022YFA1503103, 2021YFA1500700), Natural Science Foundation of China (22033002, 92261112), Natural Science Foundation of Jiangsu Province, Major Project (BK20222007), the NOMAD Center of Excellence (European Union's Horizon 2020 research and innovation program, grant agreement Nº 951786), the ERC Advanced Grant TEC1p (European Research Council, grant agreement Nº 740233), and the project FAIRmat (FAIR Data Infrastructure for Condensed-Matter Physics and the Chemical Physics of Solids, German Research Foundation, project Nº 460197019). \\

\section*{Author contributions}
SL performed all calculations and the analysis, supervised by LMG and CC. LMG, CC, and MS designed the project. All authors contributed in writing the manuscript.

\section*{Competing interests}
The authors declare no competing interests.


\section*{Appendix: methods}
\subsection{Data Set} 
The first-principle calculations for liquid silicon, solid phases of silicon, diamond vacancy, hexagonal diamond interstitial and diamond surface Si(100) are performed using DFT-GGA with the PW91 exchange-correlation functional\cite{perdew1992atoms}, implemented in the FHI-aims code using ‘light’ basis sets \cite{blum2009ab}. To simulate the structure of liquid silicon, we use constant-pressure ($P = 0$~GPa) AIMD. The calculations are performed at liquid silicon's experimental density (2.57 g/cm$^3$).  A ${2 \times 2 \times 2}$ supercell of the eight-atom diamond cubic cell (64 atoms) is heated from $T = 0$~K to $T = 5\,000$~K for rapid melting over $5\,000$ 1 fs time steps and then equilibrated at $T = 2\,000$~K for $5\,000$ 1 fs time steps. The other four AIMD simulations at different densities but the same temperature are performed, which are randomly picked up around the experimental density (namely, 1.89, 2.21, 2.94, and 3.36 g/cm$^3$). This means that there are five AIMD trajectories of liquid silicon, each of which has $5\,000$ configurations. Their RDF is calculated and averaged using the tools included in the Atomic Simulation Environment package\cite{larsen2017atomic}. 

The mixed training set was constructed by randomly sampling the same number of configurations from each of the 5 AIMD trajectories. For instance, one configuration was randomly selected from one AIMD trajectory and a total of five configurations were gathered to form the training set. In each experiment, a validation set of $N=500$ configurations was used for hyperparameter optimization of the ENNE-IP, and the performance was evaluated on a hold-out test set of $1\,000$ configurations.

To study the uncertainty of liquid silicon at extra temperatures, 5 AIMD simulations were performed at temperatures of $1\,600$~K, $1\,800$~K, $2\,200$~K, $2\,400$~K, and $2\,600$~K. These temperatures are between the experimental melting and boiling points of silicon ($1\,687$~K and $2\,628$~K, respectively) at standard pressure. The simulations were run for $5\,000$ 1 fs time steps with a constant experimental density of liquid silicon, and the initial configuration was taken from the last configuration of the melting trajectory at the same density. A test set consisting of all $5\,000$ configurations from each AIMD trajectory was created for each temperature.

For the potential energy surface, we deformed the lattice to the target volume and relaxed it with respect to the unit cell shape and atomic position, while approximately constraining the volume and also constraining the symmetry to remain that of the initial structure. For diamond vacancy and hexagonal diamond interstitial, the formation energy of ${2 \times 2 \times 2}$, ${3 \times 3 \times 3}$, ${4 \times 4 \times 4}$ and ${5 \times 5 \times 5}$ supercell of the eight-atom diamond cubic cell are calculated. The formation energies $E_{f}$ were obtained from the total energies of the supercells with a defect,
$$E_{f}=E_{d}-E_{bulk}-{n_d{\mu}}$$
where $E_{d}$ is the total energy of the defected supercell and $E_{bulk}$ is the total energy of the perfect supercell, which we calculate for supercells of the same size as used in the runs with the defects. ${\mu}$ is the chemical potential of silicon, which we take as the total energy of silicon. $n_d$ gives the number of silicon atoms that are added ($n_d$ positive) or removed ($n_d$ negative). For diamond surface Si(100), a ${(2\times 1)}$ supercell contains a slab of silicon $L$ layers thick (${L=2,3,...,10}$) and the vacuum thickness between adjacent slabs along the [001] direction is set at 15~\AA. The three outermost layers of the slab are allowed to relax, while the atoms of the other layers are kept in their bulk positions. The surface energy $\gamma$ is calculated from the following expression,
$$\gamma=\frac{E_{slab}-{n_{s}}\times{E_{bulk}}}{2A}$$
where $E_{slab}$ is the total energy of the surface slab. $n_{s}$ is the number of atoms in the surface slab. $E_{bulk}$ is the bulk energy per atom. $A$ is a measure of the surface area. Here, we define $A$ as the number of atoms on the surface (equal to 2) for a better comparison swith defect phases. For a slab, we have two surfaces and they are of the same type, which is reflected by the number 2 in the denominator. 

\subsection{Hyperparameter Optimization} 
The NequIP framework builds-in equivariance by imposing that the features at every hidden layer of the neural network be direct sums of irreducible representations. In 3-dimension space, these can be interpreted as spherical harmonics. For the practical construction of networks, we selected the highest angular frequency or degree for hidden layer features. We trained four different, equal-size, neural networks with $lmax$ of 0, 1, 2 and 3 on the training sets with different sizes of 500, $1\,000$, $2\,000$, $3\,000$, $4\,000$, and $5\,000$ configurations. Learning curves for this experiment are presented in Fig. S2 in the ESI, As expected, the training errors decrease with more training data, and the equivariant networks with $lmax$ = 3 significantly outperform the invariant networks with $lmax$ = 0 and equivariant networks with $lmax$ = 1, 2 for all data set sizes as measured by the MAE of force components. Interestingly, good calibration in terms of the MAE was obtained with relatively small training sets and the MAE does not vary significantly when adding more data in the neural networks with $lmax$ = 3. Consequently, we found that $5\,000$ configurations are sufficient to build an accurate NN interatomic potential for liquid silicon.

Then, another two key hyperparameters, the multiplicity of the features and the cutoff radius, are optimized. The results show that the optimal feature dimension is 16 and the optimal cutoff radius distance is 5.0 \AA~for liquid silicon. Finally, networks with optimal hyperparameters are trained using a loss function based on a weighted sum of energy and force loss terms (the weighting of energy and force is 1 and 100 respectively). We normalize the target energies by subtracting the mean potential energy over the training set and scale both the target energies and target force components by the root mean square of the force components over the training set. All models are trained using 4 interaction blocks, a learning rate of 0.01 and a batch size of 1. The invariant radial networks act on a trainable Bessel basis of size 8 and are implemented with two hidden layers of 32 neurons with SiLU nonlinearities between them. Considering the uncertainty of the models, we trained 5 NN models, which have the same architecture but are initialized independently, in all experiments of hyperparameter optimization.

\subsection{Uncertainty and actual error}
Our approach to estimating the uncertainty of the prediction of NNs relies on the prediction of an ensemble model consisting of several instances. In detail, the AIMD database of liquid silicon is randomly sub-sampled into 8 smaller training sets. Then the ensemble model is trained over a given training set by changing each time the initial condition of each NN. Here, the members of the ensemble ideally have the same architecture but are initialized independently via the input of different seeds. For the final prediction of the ensemble, we computed the mean of the  models' predictions $\overline{P_{i}}$, 
$$ \overline{P_{i}} = \frac{1}{M}\sum_{m}^{M} P_{i}^{m} $$
and for the uncertainty estimate, we computed the standard deviation $ \sigma_{i} $ of the set of predictions, 
$$ \sigma_{i} = \sqrt{\frac{1}{M}\sum_{m}^{M} (P_{i}^{m} - \overline{P_{i}})^2} $$
where the subscript $i$ gives configurations in the test set (atomic indices for the force), and $ m $ iterates over models of the ensemble. $ M $ is the number of neural networks in one ensemble. $ P $ is the target property, including the force component, the total energy of liquid silicon, the total energy of solid phase, the formation energy of defect silicon and the surface energy of the surface Si(100). 
The actual error of the ensemble $\overline \epsilon_{i} $ is calculated as follow,
$$ \overline{\epsilon_{i}} = \frac{1}{M}\sum_{m}^{M} (P_{i}^{m}- P_{i}^{DFT})$$

\end{document}